\documentstyle[eqsecnum,aps]{revtex}

\begin{document}
\draft
%\preprint{hep-th/9905001}
\title{Massive torsion modes from\\
Adler-Bell-Jackiw and scaling anomalies }
\author{Lay Nam Chang\cite{byline1}}
\address{
Department of Physics, Virginia Tech.\\
Blacksburg, VA24061-0435, USA. }
\author{Chopin Soo\cite{byline2}\\
}
\address{Department of Physics,\\
National Cheng Kung University\\
Tainan 70101, Taiwan. }

%\date{\today}
\maketitle
\begin{abstract}
Regularization of quantum field theories introduces a mass scale
which breaks axial rotational and scaling invariances. We
demonstrate from first principles that axial torsion and torsion
trace modes have non-transverse vacuum polarization tensors, and
become massive as a result.
%%%!!!
The underlying reasons are similar to those responsible for the
Adler-Bell-Jackiw (ABJ) and scaling anomalies.
%%%!!!
Since these are the only torsion components that can couple
minimally to spin $\frac{1}{2}$ particles, the anomalous
generation of masses for these modes, naturally of the order of
the regulator scale, may help to explain why torsion and its
associated effects, including CPT violation in chiral gravity,
have so far escaped detection. As a simpler manifestation of the
reasons underpinning the ABJ anomaly than triangle diagrams, the
vacuum polarization demonstration is also pedagogically useful.

\end{abstract}

\widetext

\section{Introduction and Overview}
\label{sec:level1}

   Torsion arises naturally in Riemann-Cartan spacetimes when
the vierbein, $e_{\mu A}$, and spin connection, $A_{\mu AB}$,
 are assumed to be independent.\footnote{In Appendix A, notations
and definitions are clarified, and a few relevant identities are
briefly introduced.} There is no compelling physical reason which
forces torsion to vanish identically. However not all components
of torsion interact naturally with matter. In minimal coupling
schemes only spinors couple to torsion, and even then only the
axial and trace modes of torsion couple to spin $\frac{1}{2}$
particles. Actually, in a hermitian theory, only the axial
torsion mode, ${\tilde A}_\mu $, interacts minimally with spin
$\frac{1}{2}$ matter.

Investigations have nevertheless uncovered that torsion can have
quite interesting quantum field theoretic effects. Recent works
have revealed that even the well-studied Adler-Bell-Jackiw (ABJ)
anomaly \cite{ABJ} receives further contributions from torsional
invariants \cite{Yajima,Chandia,cpabj}. Moreover, because of the
axial vector coupling,
%%%!!!
vacuum polarization diagrams with two external axial torsion
vertices are not transverse, with the divergence being controlled
by the Nieh-Yan term \cite{Nieh-Yan}. This breakdown in
transversality occurs in addition to that manifested by the ABJ
triangle diagrams that give rise to a term quadratic in the
curvatures.
%%the ABJ anomaly in four dimensions {\it is already manifest
%%at the level of vacuum polarization diagrams} with two external
%%axial torsion vertices \cite{cpabj},
%%in the form of the Nieh-Yan term \cite{Nieh-Yan}.
%%This is novel since the usual quadratic in curvature ABJ anomaly terms
%%come from triangle diagrams.
%%%!!

%%%Potential input
%%%!!!
%%\input mass_00s

A striking consequence of non-transversality in the polarization
tensor is the generation of mass.  We may think of the axial
torsion mode as an ``axial torsion photon" \cite{cpabj}, coupled
to a current whose conservation has been compromised because of
anomaly considerations.  The associated  ``gauge" invariance, the
$\gamma^5$ rotational symmetry,  is therefore broken, and a mass
results. In this paper, we describe explicitly how this
phenomenon takes place.

It should be noted that the breakdown in current conservation
poses no consistency problems.  The reason is that the axial
torsion modes are not gauge field modes, and are not responsible
for any local symmetries.  As shown in Appendix A, the axial
torsion field is itself a combination of vierbein and spin
connections and transforms covariantly under Lorentz and
diffeomorphism transformations.  The presence of a mass term
therefore does not jeopardize consistency of the quantum field
theory.  The computations to be described below are carried out
using a regularization that preserves local Lorentz and
diffeomorphism symmetries, and the internal symmetries of the
standard model, so all these invariances and their associated
Green functions satisfy the usual Ward identities.

This paper will also examine one characteristic aspect of the
standard model and its implications in gravity.  The standard
model incorporates maximal parity and charge conjugation
non-conservation by assigning left- and right- handed fermions to
different representations of the internal gauge group.  In 4D, it
is always possible to re-write a right-handed fermion as a
left-handed field with conjugate properties.  In this respect,
therefore, the standard model can be defined using only fermion
fields of one chirality.

In the presence of gravity, fields of one chirality are coupled
to one set of spin connections, and those of the opposite
chirality are coupled to their conjugates.  In Appendix A, we
show that the result in using only left-handed fermion fields is
a minimal coupling to the axial torsion and torsion trace fields
via the combination $J^\mu C_\mu \equiv J^\mu (i B_\mu +
A_\mu)$.   Here $J_\mu$ is the total singlet current over all
left-handed fermion fields, ${\tilde A}_\mu = 4e{A}_\mu$, the
axial torsion field, while $B_\mu$ is the trace of the torsion
field.  The first term in $C_\mu$ is anti-hermitian relative to
the second, and hermitizing the action would eliminate it
completely. But doing so requires bringing in right-handed
conjugate Weyl fields, which in turn are coupled to the self-dual
or right-handed spin connection fields. These fields are distinct
components of the spin connection, and it has already been shown
in Ref. \cite{Ash} that general relativity field equations can be
reproduced without reference to them.  They are therefore
irrelevant in determining the equations of motion. The system can
be defined by an action that is dependent upon only fields of one
chirality, thereby extending this characteristic feature of the
standard model to cover gravity interactions as
well\cite{cps,cpt}. In this scheme of things, the appearance of
$iB_\mu$ is entirely natural.

Such an action preserves holomorphy in the chiral fields.  In the
regularization scheme adopted in this paper\cite{invariant}, this
property is maintained, and both torsional modes must therefore
have the same mass.  The generation of mass for the $B$ field is
not self-evident, since it has a vector coupling to the fermion
fields rather than an axial vector coupling.  Based upon our
experience with QED, we might have argued that the associated
quanta remain massless.  The difference arises because of the
anti-hermitian coupling.  The operator appearing in the
regularized integrals involves the positive definite form in
Euclidean space
$i{D\kern-0.15em\raise0.17ex\llap{/}\kern0.15em\relax}
(i{D\kern-0.15em\raise0.17ex\llap{/}\kern0.15em\relax})^\dagger$.
Now
$(i{D\kern-0.15em\raise0.17ex\llap{/}\kern0.15em\relax})^\dagger$
differs from
$i{D\kern-0.15em\raise0.17ex\llap{/}\kern0.15em\relax}$, because
of the anti-hermitian nature of the coupling. For conventional
internal gauge fields, the two forms of the covariant derivatives
are the same, and lead to hermitian Dirac operators. We detail in
Section II how this difference causes the vacuum polarization
tensor for $B_\mu$ to be different from the usual result for
gauge theories and how $B_\mu$ acquires an anomalously generated
mass. The coefficient is equal to a similar piece involving the
axial torsion field.

We begin our discussion by showing explicitly below how the axial
torsion field can acquire a mass as a result of the
considerations outlined above, utilizing two methods of
computation.  Next, in Section II, we generalize the discussion
to cover the completely chiral action in a general curved
spacetime manifold with torsion.  We show how both the axial and
trace torsion components must acquire masses at the same time
through the covariant combination $C_\mu = iB_\mu + A_\mu$. We
end in section III with further remarks on the distinction
between gauged and ungauged symmetries, Lorentz invariance and
Abelian couplings, and also CPT violation in chiral gravity.
Appendix A contains a summary of the notations used in the
paper.  Appendix B contains details on the computation of the
polarization tensor for internal gauge fields within the
completely chiral context employed in the paper. The results are
not new\cite{Aoki,Okuyama}; without spontaneous symmetry
breaking, the tensors are all purely transverse. We have included
the details for a number of reasons. First, this invariant
regularization differs from conventional schemes in utilizing an
infinite tower of massive Pauli-Villars regulators. This is
necessary for Weyl fermions when the gauge coupling
representation $T^a$ is complex, as it is in the standard model.
Second, the doubling in the regulators typical of this method is
done in internal $T^a$ gauge space rather than by including both
left- and right- handed chiral fermions.
%The latter means is the normal procedure for Pauli-Villars
%regularization\cite{PeskinBook}.
This difference is not really significant for the vacuum
polarization of internal gauge fields in flat spacetime, but for
gravitational fields doubling in the internal space allows us to
avoid the introduction of right-handed fermions and right-handed
spin connections and thereby preserve the chirality of the
gravitational interaction even at the level of regularization.
Thus even in teleparallel spacetimes with flat vierbein, the
chirality of the torsion coupling of Weyl fermions to $C_\mu =
iB_\mu + A_\mu$ will be ensured at the quantum field theoretic
level. Appendix B also supplies the check that computations with
doubling in internal gauge space can be carried out consistently.
Finally, and more importantly, the explicit intermediate steps
allow us to compare and contrast with the results of the
polarization of the torsion modes in Section II.

%%%!!!

%\end{document}

\subsection{Vacuum polarization and the ABJ
%%%!!!
Current}
%%anomaly
%%%!!!

Consider the action of a bispinor theory in teleparallel
spacetimes with flat vierbein $e^A_{\mu} = \delta^A_\mu$ but with
nontrivial axial torsion coupling
\begin{equation}
S = - \frac{1}{2}\int d^4x \, e{\overline \Psi}\gamma^\mu
(i\partial_\mu + \frac{1}{4e}{\tilde A}_\mu\gamma^5 )\Psi + H.
c.  \label{flataction}
\end{equation}
%%%!!!
The coupling is through the axial ABJ current
%%%!!!
$J^{5\mu} ={\overline\Psi}\gamma^\mu \gamma^5 \Psi$
%%%!!!
which has expectation value
%%%!!!
\begin{equation}
\langle J^{5\mu}(x) \rangle =-\langle {{\delta S}\over {\delta
{A}_\mu(x)}}\rangle = -\lim_{x \rightarrow y} {\rm
Tr}\{\gamma^\mu \gamma^5 {1 \over
{(i{\partial\kern-0.15em\raise0.17ex\llap{/}\kern0.15em\relax} +
{A\kern-0.15em\raise0.17ex\llap{/}\kern0.15em\relax}\gamma^5)}}\delta(x-y)\},
\end{equation}
where we have defined $A_\mu \equiv \frac{1}{4e}{\tilde A}_\mu$
for convenience. The corresponding vacuum polarization tensor,
$\Pi^{\mu \nu}$, is defined by the Fourier transform
\begin{equation}
\left.{{\delta \langle J^{5\mu}(x) \rangle}\over {\delta
A_\nu(y)}}\right|_{A_{\alpha}=0} = \int {{d^4k} \over {(2\pi)^4}}
\Pi^{\mu\nu}(k) e^{ik.(x-y)}, \label{e3}
\end{equation}
from which
\begin{equation}
\Pi^{\mu\nu}(k) \propto \int d^4p \, {\rm Tr}\{\gamma^\mu\gamma^5
{1 \over {{p\kern-0.15em\raise0.17ex\llap{/}\kern0.15em\relax} +
{k\kern-0.15em\raise0.17ex\llap{/}\kern0.15em\relax}}}
\gamma^\nu\gamma^5 {1\over
{{p\kern-0.15em\raise0.17ex\llap{/}\kern0.15em\relax}}}\}.
\end{equation}
Had this expression been well defined, it would have been no more
than
\begin{equation}
\Pi^{\mu\nu}(k) \propto \int d^4p  \, {\rm Tr}\{\gamma^\mu {1
\over {{p\kern-0.15em\raise0.17ex\llap{/}\kern0.15em\relax} +
{k\kern-0.15em\raise0.17ex\llap{/}\kern0.15em\relax}}} \gamma^\nu
{1\over {{p\kern-0.15em\raise0.17ex\llap{/}\kern0.15em\relax}}}\}
\end{equation}
since the two $\gamma^5$'s cancel out in the trace, and we would
have obtained the usual vacuum polarization amplitude for which
we do not expect a longitudinal component. But it is not, for the
integration over fermion loop momentum diverges.
%%%!!!
We will need a regularization scheme, Pauli-Villars for instance,
to tame this divergence before performing any Dirac algebra. Any
gauge-invariant scheme however will compromise symmetry generated
by the axial current; this is the essence of the ABJ anomaly.
For example, in the scheme to be used in this paper, the
regulator fields carry masses $\{m_n\}$.
%%{\it Regularization} (using for instance Pauli-Villars regulators with
%%masses $\{m_n\}$) {\it introduces a subtlety} which is the very
%%basis of the ABJ anomaly.
%%%!!!
Summing over the propagators for all the fields, including the
massive regulators, results in
\begin{equation}
\Pi^{\mu\nu}(k) \propto \sum_n C_n \int d^4p \, {\rm
Tr}\{\gamma^\mu\gamma^5 {1 \over
{({p\kern-0.15em\raise0.17ex\llap{/}\kern0.15em\relax} +
{k\kern-0.15em\raise0.17ex\llap{/}\kern0.15em\relax}) + i m_n }}
\gamma^\nu \gamma^5 {1 \over
{{p\kern-0.15em\raise0.17ex\llap{/}\kern0.15em\relax} + i m_n}}\},
\end{equation}
with $C_n = \pm 1$, depending on whether the regulators are
anticommuting or commuting. (For the original fermion multiplet,
$C_0 =1$ and $m_0= 0$. We assume analytic continuation to
Euclidean Green functions.) By moving the second $\gamma^5$ to the
left to cancel out the first, and bearing in mind that $\gamma^5$
anticommutes with the Dirac matrices, we observe that $m_n$ {\it
changes its relative sign} with respect to
$({p\kern-0.15em\raise0.17ex\llap{/}\kern0.15em\relax} +
{k\kern-0.15em\raise0.17ex\llap{/}\kern0.15em\relax})$ in the
denominator. Consequently,
\begin{equation}
\Pi^{\mu\nu}(k) \propto \sum_n C_n \int d^4p \, {\rm
Tr}\{\gamma^\mu{1 \over
{({p\kern-0.15em\raise0.17ex\llap{/}\kern0.15em\relax} +
{k\kern-0.15em\raise0.17ex\llap{/}\kern0.15em\relax}) -i m_n }}
\gamma^\nu {1 \over
{{p\kern-0.15em\raise0.17ex\llap{/}\kern0.15em\relax} +i m_n}}\}.
\label{e1}
\end{equation}
The integrals over the loop momentum will be well defined for a
suitable set $\{C_n, m_n\}$ which satisfies the Pauli-Villars
conditions. In Section II an explicit set of $\{C_n, m_n \}$
shall be presented for the Weyl theory. It is also applicable to
the bispinor theory here, consistently yielding a polarization
magnitude which is twice that of the single Weyl fermion. The
important point is that if we had started with a vector (instead
of the axial vector) coupling, the result for Eq.\ (\ref{e1})
would have been
\begin{equation}
\Pi^{\mu\nu}(k) \propto \sum_n C_n \int d^4p \, {\rm
Tr}\{\gamma^\mu {1 \over
{({p\kern-0.15em\raise0.17ex\llap{/}\kern0.15em\relax} +
{k\kern-0.15em\raise0.17ex\llap{/}\kern0.15em\relax}) +i m_n }}
\gamma^\nu {1 \over {{p\kern-0.15em\raise0.17ex\llap{/}
\kern0.15em\relax} +i m_n}}\} \label{e2}
\end{equation}
instead.
%%%!!!
This integral would have produced a transverse polarization
tensor.  As it is, by rewriting one of the propagators in Eq.\
(\ref{e1}) as
%%%!!!
\begin{equation}
{1 \over {({p\kern-0.15em\raise0.17ex\llap{/}\kern0.15em\relax} +
{k\kern-0.15em\raise0.17ex\llap{/}\kern0.15em\relax}) -i m_n}} =
{1 \over {({p\kern-0.15em\raise0.17ex\llap{/}\kern0.15em\relax} +
{k\kern-0.15em\raise0.17ex\llap{/}\kern0.15em\relax}) +i m_n}} -
{2i m_n \over {(p + k)^2 + m^2_n}}.\label{two}
\end{equation}
%%%!!!
we obtain two terms, the first of which is identical to the
integral in Eq.\ (\ref{e2}).  However,
%%and substituting this back into Eq.\ (\ref{e1}) gives
%%the first contribution
%%contribution precisely as in Eq.\ (\ref{e2}), but
there is now an additional anomalous term which is given by
\begin{equation}
\sum_n C_n \int d^4p \, {\rm Tr}\{\gamma^\mu {2i m_n \over {(p +
k)^2 + m^2_n }} \gamma^\nu
{({p\kern-0.15em\raise0.17ex\llap{/}\kern0.15em\relax} -i m_n)
\over {p^2 + m^2_n}}\} = 8g^{\mu\nu} \sum_n C_n \int d^4p
\,{{m^2_n} \over {[(p + k)^2 + m^2_n][p^2 + m^2_n] }}.
\end{equation}
This non-transverse part of the vacuum polarization tensor, above
and beyond the usual transverse $(k^\mu k^\nu - g^{\mu\nu}
k^2)\Pi(k^2)$ term from Eq.\ (\ref{e2}), is clearly generated
anomalously through massive regulators which break the $\gamma^5$
symmetry of the classical action. Both the axial vector coupling
and the presence of nontrvial $m_n$'s from the regularization are
required for the argument to go through. From
\begin{equation}
\Pi^{\mu\nu} = (k^\mu k^\nu - g^{\mu\nu} k^2)\Pi + g^{\mu\nu}\Pi'
\end{equation}
and the divergence of Eq.\ (\ref{e3}), we deduce that
\begin{equation}
k_\mu \Pi^{\mu\nu} \propto k^\nu \quad {\rm and} \quad \langle
\partial_\mu J^{5\mu}\rangle_{\rm Reg} \propto \partial_\mu
{\tilde A}^\mu \neq 0
\end{equation}
at the level of vacuum polarization diagrams. The anomalous
$g^{\mu\nu}\Pi'$ contribution in the vacuum polarization tensor
also implies that besides the usual $F_{\mu\nu}F^{\mu\nu}$ piece
required for the transverse polarization, a mass counter term of
the form $A_\mu A^\mu$ is also needed for the non-zero
longitudinal component in the effective action. $A_\mu$ becomes
massive as a result.

%------------------------------------------------------------
In vector QED, curbing divergences by naive momentum truncation
%%%!!!
also
%%%!!!
results in a non-transverse photon polarization tensor
\cite{PeskinBook}.
%%%!!
But this
%%%!!!
%%This
apparent breakdown of gauge invariance is an artifact of symmetry
breaking ``regularization" which can be removed altogether by
proper gauge-invariant regularization schemes. One may therefore
suspect that, in the axial-vector coupling of torsion to
fermions, the non-transverse polarization exhibited here is
likewise
%%merely
the consequence of a ``fake anomaly" resulting from ``improper
regularization" which breaks the symmetry of axial rotations.
%%%!!!
That this is not the case is guaranteed by the ABJ anomaly.  The
anomaly assures us that there are no regularization schemes that
preserve singlet axial rotations, and at the same time respect all
of the local symmetries, such as Lorentz and other gauge
symmetries, which are present.
%%On the contrary,
The non-transverse polarization for axial torsion can thus be
regarded as another manifestation of this phenomenon.
%%%!!!
As discussed in Appendix A, $A_\mu$ transforms covariantly under
diffeomorphisms, and is Lorentz invariant. The counter term
necessary to compensate for a non-transverse polarization tensor,
$A_\mu A^\mu$, is therefore completely consistent with local
Lorentz and diffeomorphism invariance. A similar term in QED
would have violated local gauge invariance.
%%ABJ anomaly
%%which cannot be disposed of; and
%%%!!!
%%%!!! Next paragraph replaced.
%%There are indeed crucial distinctions
%%between gauged vector QED coupling and axial-vector coupling of
%%torsion to the ABJ current.
%%%!!!
%%As
%%evidenced in
%%the computations above (and also later on in Section II), a single
%%invariant regularization scheme produces at the same time transverse
%%polarization for gauged photons but non-transverse polarization
%%for the torsion fields (which are Lorentz invariant composites).
%%In usual practice ``fake anomalies" can be removed by adding
%%gauge-noninvariant counterterms but $A_\mu A^\mu$ is in fact locally
%%Lorentz
%%and gauge invariant. Moreover, there are really no $\gamma^5$invariant
%%regularizations with the correct continuum physical spectrum which one
%%can hope to introduce. Other considerations will further demonstrate
%%that
%%the $A_\mu A^\mu$ term must be present and Eqs. (1.11) and (1.12) are
%%indeed correct.

%%%!!!
Before leaving this topic, it is instructive to exhibit and
confirm the same effect using an alternative regularization
method. For bispinors, we may write the effective action in
curved space as
%%%!!!
\begin{eqnarray}
\Gamma_{\rm eff.} &=& -i {\rm Tr}\ln [e^{1\over 2}i
{\Delta\kern-0.15em\raise0.17ex\llap{/}\kern0.15em\relax}
e^{-{1\over 2 }}], \cr
\nonumber\\
i{\Delta\kern-0.15em\raise0.17ex\llap{/}\kern0.15em\relax} &=&
\gamma^\mu(i\partial_\mu + {i\over 2}\omega_{\mu AB}\sigma^{AB} +
A_\mu \gamma^5).
\end{eqnarray}
%%%!!!
With heat kernel regularization and the Schwinger-DeWitt
expansion\cite{DeWitt}, the ABJ anomaly with Euclidean signature
has been demonstrated to take the form\cite{Yajima}
\begin{eqnarray}
\langle \partial_\mu J^{5\mu}\rangle_{\rm Reg.} &=& 2i\lim_{t
\rightarrow 0}\lim_{x \rightarrow x'}{1\over{(4\pi t)}^2} {\rm
Tr}\langle x'|\gamma^5
\exp[-t(i{\Delta\kern-0.15em\raise0.17ex\llap{/}\kern0.15em\relax})^2]
|x\rangle \cr
\nonumber\\
&=& 2i\lim_{t \rightarrow 0}{1\over{(4\pi t)}^2} {\rm
Tr}[\sum^\infty_{n=0} e\gamma^5 a_n t^n].
\end{eqnarray}
Furthermore, it is known that the traces of the coefficients
$a_0, a_1$ and $a_2$ contribute to the divergent part of the
effective action. For instance, $a_0 = I$ leads to the
renormalization of the cosmological constant. We focus on the
relevant coefficient $a_1$ which, in our notation,
%%%!!!
%%I have eliminated the subscript $\omega$ in $R$ to avoid unnecessary
%%confusion
%%on the part of the reader%%
is $a_1 = -\frac{1}{12}R -2A_\mu A^\mu - \gamma^5
e^{-1}\partial_\mu(eA^\mu)$. Clearly ${\rm Tr}(e a_1)$ (and thus
the effective action) contains both the familiar Einstein-Hilbert
term, $e R$, as well as the $e A_\mu A^\mu$ axial torsion
contribution which we also found to be required by the vacuum
polarization computations above. Moreover, it follows from
Eq.(1.13) that the ABJ anomaly is
\begin{equation}
\langle \partial_\mu J^{5\mu}\rangle_{\rm Reg} = -{{2i}\over
{(4\pi)^2 t}}\partial_\mu \tilde A^\mu + {{2i}\over
{(4\pi)^2}}{\rm Tr}(e\gamma^5 a_2).
\end{equation}
${\rm Tr}(e\gamma^5 a_2)$ is the more familiar regulator scale
independent part of the ABJ anomaly. But there is also the
$t$-dependent first term ($t$ has the physical dimension of
inverse regulator mass squared). This is precisely the Nieh-Yan
contribution to the ABJ anomaly.
%%%!!!
The linear dependence on $A_\mu$ shows that
%%for which
the vacuum polarization is indeed the correct Feynman diagram
process to consider for this purpose\cite{cpabj}. Therefore the
non-transversality of the polarization tensor is not only
genuine, but is in fact necessary to understand the origin of the
Nieh-Yan contribution to the ABJ anomaly in perturbation theory.

\section{Weyl fermions and Vacuum Polarization tensor of torsion fields}

%\begin{equation}
%\int d^4p = \pi^2\int^\infty_0 p^2 dp^2
%\end{equation}

The classical Weyl action for a left-handed fermion multiplet in a
general curved spacetime is
\begin{equation}
S_L = -\int d^4x\, e{\overline\Psi}_L
i{D\kern-0.15em\raise0.17ex\llap{/}\kern0.15em\relax}\Psi_L,
\end{equation}
with $i{D\kern-0.15em\raise0.17ex\llap{/}\kern0.15em\relax} =
\gamma^\mu(i\partial_\mu + \frac{i}{2}A_{\mu AB}\sigma^{AB} +
W_{\mu a}{T}^{a})$. It is known that if the representation of the
internal gauge field $T^a$ is perturbatively anomaly-free $({\rm
Tr}(T^a) ={\rm Tr}(T^a\{T^b,T^c\}) = 0)$, then all fermionic loops
in background gauge and gravitational fields of the theory can be
regularized in an explicitly gauge, Lorentz and diffeomorphism
invariant manner through an infinite tower of Pauli-Villars
regulators which are doubled in the internal space (see Ref.
\cite{invariant} for further details.) This generalizes the
invariant scheme first introduced by Frolov and
Slavnov\cite{Slavnov}. Specifically, to form general invariant
masses, the internal space is doubled from $T^a$ to
\begin{equation}
{\cal T}^a =\left(\matrix{(-T^a)^* &0\cr 0&T^a}\right);
\end{equation}
and the original fermion multiplet is projected as $\Psi_{L} =
\frac{1}{2}(1-\sigma^3)\Psi_{L}$, where
\begin{equation}
\sigma^3 =\left(\matrix{1_d &0\cr 0& -1_d}\right),
\end{equation}
and $d$ is the number of Weyl fermions in the $\Psi_{L}$
multiplet. Written in full, the regularized action is
\begin{eqnarray}
{\cal S}_{L_{\rm
Reg}}=-{\int}d^4xe[&{\sum_{r=0,2,...}}&\{{\overline\Psi}_{L_r}
i{{D\kern-0.15em\raise0.17ex\llap{/}\kern0.15em\relax}}\Psi_{L_r}
+ {1\over 2}m_r(\Psi^T_{L_r}\sigma^1C_4\Psi_{L_r} +
{\overline\Psi}_{L_r}\sigma^1C^\dagger_4{\overline
\Psi}^T_{L_r})\}\cr
-&{\sum_{s=1,3,...}}&\{{{\overline\Phi}_L}_s\sigma^3i
{{D\kern-0.15em\raise0.17ex\llap{/}\kern0.15em\relax}}\Phi_{L_s} +
{1\over 2}m_s(\Phi^T_{L_s}\sigma^1\sigma^3C_4\Phi_{L_s} +
{\overline\Phi}_{L_s}
C^\dagger_4\sigma^3\sigma^1{\overline\Phi}^T_{L_s})\}].\label{Lreg}
\end{eqnarray}
The sums are over all even natural numbers for the anticommuting
and over all odd natural numbers for the commuting fields. With
the exception of the original undoubled and massless ($m_0 = 0$)
multiplet (written as $\Psi_{L_0} \equiv \Psi_L$), all other
anticommuting $\Psi_{L_r}$ and commuting $\Phi_{L_s}$ multiplets
are generalized Pauli-Villars regulator fields, doubled in
internal space, and endowed with Majorana masses, which we take
for definiteness to satisfy $m_n = n\Lambda$. $C_4$ is the charge
conjugation matrix in four dimensions, and in the covariant
derivative all fields are now coupled to $W_{\mu a}{\cal T}^a$.
We emphasize that because all the multiplets are left-handed,
there are no couplings to the right-handed spin connection which
neither needs nor should be introduced for a truly Weyl theory.
The regularization therefore preserves the chirality of the
theory with respect to the gravitational interaction. It can be
shown\cite{invariant} the net effect of the regularization is to
replace the $\frac{1}{2}(1-\sigma^3)$ projection of the bare
currents by
$\frac{1}{2}(f({{{D\kern-0.15em\raise0.17ex\llap{/}\kern0.15em\relax}
{{D\kern-0.15em\raise0.17ex\llap{/}\kern0.15em\relax}^\dagger}}
\over{\Lambda^2}})-\sigma^3)$ where $f$ is the regulator function,
\begin{equation}
f({D\kern-0.15em\raise0.17ex\llap{/}\kern0.15em\relax}
{{D\kern-0.15em\raise0.17ex\llap{/}\kern0.15em\relax}}^\dagger/\Lambda^2)
=\sum_n C_n{{i{D\kern-0.15em\raise0.17ex\llap{/}\kern0.15em\relax}
({i{D\kern-0.15em\raise0.17ex\llap{/}\kern0.15em\relax}})^\dagger}
\over{[i{D\kern-0.15em\raise0.17ex\llap{/}\kern0.15em\relax}
{(i{D\kern-0.15em\raise0.17ex\llap{/}\kern0.15em\relax})^\dagger}
+ m^2_n]}} =\sum^{\infty}_{n= -\infty} {{(-1)^n
i{D\kern-0.15em\raise0.17ex\llap{/}\kern0.15em\relax}
({i{D\kern-0.15em\raise0.17ex\llap{/}\kern0.15em\relax}})^\dagger}
\over{[i{D\kern-0.15em\raise0.17ex\llap{/}\kern0.15em\relax}
{(i{D\kern-0.15em\raise0.17ex\llap{/}\kern0.15em\relax})^\dagger}
+ n^2\Lambda^2]}}.
\end{equation}

To concentrate on the vacuum polarization of the torsion fields,
we specialize to $W_{\mu a} =0$, and flat vierbein $e_{\mu A}
=\eta_{\mu A}$ but retain nontrivial torsion couplings. To wit,
the Weyl action reduces to
\begin{equation}
S_L= \int d^4x [-e{\overline\Psi}_L\gamma^\mu i\partial_\mu\Psi_L
+ C_\mu{\overline\Psi}_L e\gamma^\mu \Psi_L], \label{Laction}
\end{equation}
with $ C_\mu = iB_\mu + \frac{1}{4e}{\tilde A}_\mu$. Note that the
torsion interaction appears exclusively in this specific
combination. Discussion in Appendix A shows that the vector
$C_\mu$ is covariant relative to diffeomorphisms, but is
otherwise invariant under local Lorentz and internal symmetry
transformations. The bare current
\begin{equation}
\langle {{\delta S_L}\over{\delta C_{\nu}(x)}}\rangle =\langle
J^\mu_L \rangle_{Bare} =\lim_{x \rightarrow y} {\rm
Tr}\{\gamma^\mu\frac{1}{2}(1-\gamma^5)
{1\over{i{D\kern-0.15em\raise0.17ex\llap{/}\kern0.15em\relax}}}
\frac{1}{2}(1-\sigma^3)\delta(x-y)\}
\end{equation}
is modified by the Pauli-Villars regulators to become
\begin{eqnarray}
\langle J^{\mu}_L(x)\rangle_{\rm Reg} &=& \lim_{x \rightarrow
y}{\rm Tr}\{\gamma^\mu \frac{1}{2}(1-\gamma^5) {1\over
{i{D\kern-0.15em\raise0.17ex\llap{/}\kern0.15em\relax}}}
\frac{1}{2}(f -\sigma^3)\delta(x-y)\}\cr
\nonumber\\
&=&\lim_{x \rightarrow y}{\rm
Tr}\{\gamma^\mu\frac{1}{2}(1-\gamma^5) (\{\frac{1}{2}\sum_n
{{(-1)^n
({i{D\kern-0.15em\raise0.17ex\llap{/}\kern0.15em\relax}})^\dagger}
\over{[i{D\kern-0.15em\raise0.17ex\llap{/}\kern0.15em\relax}
{(i{D\kern-0.15em\raise0.17ex\llap{/}\kern0.15em\relax})^\dagger}
+ m^2_n]}}\}
-{1\over{2i{D\kern-0.15em\raise0.17ex\llap{/}\kern0.15em\relax}}}\sigma^3)
\delta(x-y)\}.
\end{eqnarray}
As demonstrated in Ref.\cite{invariant}, the $\sigma^3$ part
vanishes automatically for fermion loops with four or less
external vertices, and hence does not contribute to vacuum
polarization diagrams.

The full curved space Dirac operator satisfies\footnote{This can
be shown with the identity $\partial_\mu \gamma^\mu +
(\partial_\mu \ln e)\gamma^\mu + \frac{1}{2}A_{\mu
AB}[\sigma^{AB},\gamma^\mu] =
2{B\kern-0.15em\raise0.17ex\llap{/}\kern0.15em\relax}$, which
follows from $\gamma^\mu = E^\mu_A \gamma^A$ and the metricity
($\nabla_\mu E^{\nu A} = \nabla_\mu e_{\nu A} = 0$) of the
GL(4,R) connection $\Gamma$ introduced in $\nabla = d + \Gamma +
A$.}
\begin{equation}
(i{D\kern-0.15em\raise0.17ex\llap{/}\kern0.15em\relax})^\dagger =
i{D\kern-0.15em\raise0.17ex\llap{/}\kern0.15em\relax} +
2i{B\kern-0.15em\raise0.17ex\llap{/}\kern0.15em\relax}.
\end{equation}
with respect to the Euclidean\footnote{In continuing from
Lorentzian $(-, +, +, +)$ to Euclidean signature $(+,+,+,+)$ our
Dirac matrices satisfy ${\gamma^\mu}^\dagger = \gamma^\mu$.} inner
product $\langle X|Y \rangle = \int d^4x \, e X^\dagger Y$.
Therefore
\begin{equation}
i{D\kern-0.15em\raise0.17ex\llap{/}\kern0.15em\relax} =
i{\partial\kern-0.15em\raise0.17ex\llap{/}\kern0.15em\relax} -
(i{B\kern-0.15em\raise0.17ex\llap{/}\kern0.15em\relax} -
\frac{1}{4e}{{\tilde
A}\kern-0.15em\raise0.17ex\llap{/}\kern0.15em\relax}
\gamma^5),\qquad
(i{D\kern-0.15em\raise0.17ex\llap{/}\kern0.15em\relax})^\dagger =
i{\partial\kern-0.15em\raise0.17ex\llap{/}\kern0.15em\relax} +
(i{B\kern-0.15em\raise0.17ex\llap{/}\kern0.15em\relax} +
\frac{1}{4e}{{\tilde
A}\kern-0.15em\raise0.17ex\llap{/}\kern0.15em\relax} \gamma^5),
\end{equation}
and the positive-definite operator which appears in the regulator
function $f$ is
\begin{equation}
i{D\kern-0.15em\raise0.17ex\llap{/}\kern0.15em\relax}
(i{D\kern-0.15em\raise0.17ex\llap{/}\kern0.15em\relax})^\dagger =
-\Box +
i{\partial\kern-0.15em\raise0.17ex\llap{/}\kern0.15em\relax}
(i{B\kern-0.15em\raise0.17ex\llap{/}\kern0.15em\relax} +
\frac{1}{4e}{{\tilde
A}\kern-0.15em\raise0.17ex\llap{/}\kern0.15em\relax}
\gamma^5)-(i{B\kern-0.15em\raise0.17ex\llap{/}\kern0.15em\relax}
- \frac{1}{4e}{{\tilde
A}\kern-0.15em\raise0.17ex\llap{/}\kern0.15em\relax}
\gamma^5)i{\partial\kern-0.15em\raise0.17ex\llap{/}\kern0.15em\relax}
-(i{B\kern-0.15em\raise0.17ex\llap{/}\kern0.15em\relax} -
\frac{1}{4e}{{\tilde
A}\kern-0.15em\raise0.17ex\llap{/}\kern0.15em\relax}
\gamma^5)(i{B\kern-0.15em\raise0.17ex\llap{/}\kern0.15em\relax} +
\frac{1}{4e}{{\tilde
A}\kern-0.15em\raise0.17ex\llap{/}\kern0.15em\relax} \gamma^5)
\end{equation}
where $\Box =\partial_\mu \partial^\mu$. In computing the vacuum
polarization $\Pi^{\mu\nu}$ defined as
\begin{equation}
\int {{d^4k}\over{(2\pi)^4}}\, \Pi^{\mu\nu}{e^{ik.(x-y)}} \equiv
\left.{{\delta \langle J^{\mu}_L(x)\rangle}\over{\delta
C_{\nu}(y)}} \right|_{C_\alpha =0},
\end{equation}
we need retain only terms linear in $\tilde A_\mu$ and $B_\mu$ in
the regularized current. Displaying only the relevant terms,
\begin{eqnarray}
\langle J^{\mu}_L(x)\rangle_{\rm Reg} =&& \lim_{x \rightarrow
y}{\rm Tr}\{\frac{1}{2}\gamma^\mu\frac{1}{2}(1-\gamma^5) \sum_n
(-1)^n [... +{{m_n}\over{(-\Box + m^2_n)}}
(i{B\kern-0.15em\raise0.17ex\llap{/}\kern0.15em\relax}
+\frac{1}{4e}{{\tilde
A}\kern-0.15em\raise0.17ex\llap{/}\kern0.15em\relax}
\gamma^5){m_n\over{(-\Box + m^2_n)}} \cr \nonumber\\
&+&{{i{\partial\kern-0.15em\raise0.17ex\llap{/}\kern0.15em\relax}}
\over{(-\Box +
m^2_n)}}(i{B\kern-0.15em\raise0.17ex\llap{/}\kern0.15em\relax} -
\frac{1}{4e}{{\tilde
A}\kern-0.15em\raise0.17ex\llap{/}\kern0.15em\relax}
\gamma^5){{i{\partial\kern-0.15em\raise0.17ex\llap{/}\kern0.15em\relax}}
\over{(-\Box + m^2_n)}} +...]\delta(x-y)\}.
\end{eqnarray}
After moving the $\gamma^5$ associated with ${\tilde A}_\mu$ to
the left, this works out to be
\begin{eqnarray}
\langle J^{\mu }_L(x)\rangle_{\rm Reg} =&& \lim_{x \rightarrow
y}{\rm Tr}\{\frac{1}{2}\gamma^\mu \frac{1}{2}(1-\gamma^5)\sum_n
(-1)^n [ ... + {{m_n}\over{(-\Box + m^2_n)}}
{C\kern-0.15em\raise0.17ex\llap{/}\kern0.15em\relax}
{m_n\over{(-\Box + m^2_n)}}\cr
\nonumber\\
&&+{{i{\partial\kern-0.15em\raise0.17ex\llap{/}\kern0.15em\relax}}
\over{(-\Box +
m^2_n)}}{C\kern-0.15em\raise0.17ex\llap{/}\kern0.15em\relax}
{{i{\partial\kern-0.15em\raise0.17ex\llap{/}\kern0.15em\relax}}
\over{(-\Box + m^2_n)}} +...]\delta(x-y)\},
\end{eqnarray}
Again $C_\mu$ is the only torsion combination that appears in the
final result. In the above expansion of the current it is crucial
to note the two displayed terms within square brackets has the
same sign. This is to be contrasted with the result for internal
gauge current in Eq.\ (\ref{B4}) of Appendix B in which a sign
difference occurs. Thus this origin of non-transverse torsion
polarization for Weyl theory is essentially the same as that for
the bispinor theory of Section IA. By comparing the intermediate
steps of Eqs.(2.8)-(2.14) with those in Appendix B, these
differences in the signs with Eq.\ (\ref{B4}) can be traced
precisely
 to the $\gamma^5$ which comes with ${\tilde A}_\mu$ and the
non-hermitian $iB_\mu$ interaction in
$i{D\kern-0.15em\raise0.17ex\llap{/}\kern0.15em\relax}$. The
relevant Feynman diagram processes for vacuum polarization can
also be read off from the above expression of the current. The
first term arises from nontrivial
${\overline\Psi}_L{\overline\Psi}_L$ and ${\Psi_L}{\Psi_L}$
propagators due to regulator Majorana masses, while the second
term is associated with $\Psi_L{\overline\Psi}_L$ propagators.
Proceeding as in Appendix B, we arrive at
\begin{equation}
\Pi^{\mu\nu} = d \sum_n (-1)^n \int
{{d^4p}\over{(2\pi^4)}}(-I^{\mu\nu n}_1 +I^{\mu\nu n}_2),
\end{equation}
where $I^{\mu\nu n}_{1,2}$ are as defined in Eqs.(B9) and (B10).
Again the crucial difference with Eq.\ (\ref{B11}) for internal
gauge fields is the negative sign multiplying $I^{\mu\nu n}_1$.
We decompose the expression
\begin{equation}
\Pi^{\mu\nu} = (A^{\mu\nu}_T + A^{\mu\nu}_L)d,
\end{equation}
into transverse and longitudinal contributions $A^{\mu\nu}_T$ and
$A^{\mu\nu}_L$. The transverse piece
\begin{equation}
A^{\mu\nu}_T ={1\over {24\pi^2}} (k^\mu k^\nu -
g^{\mu\nu}k^2)[\ln ({{k^2}\over{\Lambda^2}}) - \frac{5}{3} + 2
\ln ({\pi \over 2})]
\end{equation}
is identical to the result obtained in Eq.\ (\ref{B19}); while the
nontrivial anomalous longitudinal component is given by
\begin{eqnarray}
A^{\mu\nu}_L &=& -2\sum^{\infty}_{n=-\infty} (-1)^n \int
{{d^4p}\over {(2\pi)^4}}\, I^{\mu\nu n}_1 \cr
\nonumber\\
&=& 4 g^{\mu\nu}\int^1_0 dz \int {{d^4p}\over {(2\pi)^4}}\,
\sum^{\infty}_{n=-\infty} (-1)^n {{n^2\Lambda^2}\over {[p^2 +
k^2z(1-z) + n^2\Lambda^2]^2}}\cr
\nonumber\\
&=& g^{\mu\nu}{\Lambda^2 \over {(4\pi^2)}} \int^1_0 dz
\int^\infty_0 (dp^2) p^2 {d \over{dp^2}}[(p^2 + s){\tilde f}(p^2
+ s)] \cr
\nonumber\\
&=& g^{\mu\nu}{\Lambda^2 \over {(4\pi^2)}}\int^1_0 dz
\{[\left.p^2(p^2 + s){\tilde f}(p^2 +s)]\right|^\infty_{p^2 =0}
-\int^\infty_0 (p^2+s){\tilde f}(p^2 + s)dp^2\}\cr
\nonumber\\
&=& -g^{\mu\nu}{\Lambda^2 \over {(2\pi^2)}}\int^1_0 dz
\int^\infty_{\sqrt s} {{\pi p^2}\over {\sinh(\pi p)}} dp.
\end{eqnarray}
${\tilde f}$ and the properties associated with it are as in Eqs.
(B12)-(B14), and $s ={{{k^2}z(1-z)}\over{\Lambda^2}}$.
%On taking the $\Lambda \rightarrow \infty$ limit of the regulator mass
%scale,
%\begin{equation}
%\int^\infty_{0} {\pi p^2\over {\sinh(\pi p)}} dp = {7\over
%\pi^2}\zeta(3).
%\end{equation}
Although the result for the longitudinal component can be written
as integrals of polylogarithmic functions, for finite $\Lambda^2
\gg k^2$, it is more enlightening to express it as a convergent
power series in terms of Bernoulli numbers and gamma functions as
\begin{equation}
A^{\mu\nu}_L = -g^{\mu\nu}[{{7\zeta(3)} \over {4\pi^4}}\Lambda^2
+ k^2 \sum_{n=0}{{(2^{2n-1} -1)n! B_{2n}
\pi^{2n-2}{\sqrt\pi}}\over {2^{2n+4}(2n)!\Gamma(n +
\frac{5}{2})}}\left({k^2 \over \Lambda^2}\right)^n].
\end{equation}
The longitudinal part of the polarization therefore diverges as
the square of the regulator mass scale.

In the effective propagator with vacuum polarization insertions,
it is known that a non-trivial longitudinal polarization causes a
shift to a physically massive pole, even if the bare propagator
is massless in the beginning (see, for instance,
Ref.\cite{PeskinBook}). Since $\Pi^{\mu\nu}$ is the Fourier
transform of $\left.{{\delta^2 \Gamma_{eff.}} \over{{\delta
C_\nu}{\delta C_\mu}}}\right|_{C=0}$, this implies that in
addition to the more familiar curvature squared counter term
$g^{\mu \alpha}g^{\nu\beta}(\partial_\mu C_\nu -\partial_\nu
C_\mu) (\partial_\alpha C_\beta -\partial_\beta C_\alpha)$
required by the logarithmic divergence of the transverse part
$A^{\mu\nu}_T$ of $\Pi^{\mu\nu}$, a counter term proportional to
$g^{\mu\nu}C_\mu C_\nu$ for the longitudinal component
$A^{\mu\nu}_L$ of $\Pi^{\mu\nu}$ is also needed in the
Lagrangian. The presence of these terms in the effective action
implies that as a result $C_\mu$ becomes massive and obeys the
Proca equation. We discuss next what implications this mass will
have on the local invariances of the action.

\section{Further remarks}

The Weyl action of Eq.\ (\ref{Laction}) would be gauge invariant
under {\it local} $\gamma^5$ and scaling transformations
\begin{equation}
\Psi_L \rightarrow \exp(i\alpha(x)\gamma^5
-\frac{3}{2}\beta(x))\Psi_L = T \Psi_L , \qquad
e{\overline\Psi}_L\gamma^\mu \rightarrow
e{\overline\Psi}_L\gamma^\mu T^{-1} ,
\end{equation}
with $T(x)=\exp[-(i\alpha +\frac{3}{2}\beta)]$ if we {\it
pretend} that $C_\mu = (iB_\mu + A_\mu)$ is a complex Abelian
gauge connection which transforms as
\begin{equation}
C_\mu \rightarrow T C_\mu T^{-1} - T i\partial_\mu T^{-1}
\end{equation}
i.e.
\begin{equation}
A_\mu \rightarrow A_\mu +\partial_\mu \alpha ,\qquad B_\mu
\rightarrow B_\mu - \frac{3}{2}\partial_\mu\beta.
\end{equation}
Note that $B_\mu$ comes with an $i$ in the complex combination
$C_\mu$ because, unlike $\gamma^5$ rotations, the group
parametrized by $\exp(-\frac{3}{2}\beta)$ is noncompact rather
than $U(1)$. However the massive regulators of Eq.\ (\ref{Lreg})
break both of these symmetries, and the current $J^\mu_L$ coupled
to $C_\mu$ is not conserved. As shown, $k_\mu \Pi^{\mu\nu} \propto
k^\nu$ at the vacuum polarization level. The full Weyl theory
exhibits no inconsistencies because these invariances are really
not gauged as local symmetries. The theory is on the other hand
diffeomorphism, and local internal gauge and Lorentz invariant,
with internal symmetries gauged by $W_{\mu a}$ and local Lorentz
invariance by the full spin connection $A_{\mu AB}$. In fact,
$C_\mu$ transforms covariantly under general coordinate
transformations, and is invariant under Lorentz and gauge
transformations. Under global $\gamma^5$ transformations, $e_{\mu
A}$ and $A_{\mu AB}$ are inert; while $e_{\mu A} \rightarrow
\exp(\beta)e_{\mu A}$ but $A_{\mu AB}$ remains unchanged under
global scaling. Hence $C_\mu$ is invariant under both. These
global transformations are however symmetries of the bare Weyl
action. As a result $\partial_\mu J^\mu_L = 0$ and the vanishing
of the trace of the energy momentum tensor, $eT^\mu\,_\mu = 0$,
hold at the classical level, but these equations are nevertheless
anomalous at the quantum level. For the Weyl theory, there is
actually an interesting relation
\begin{equation}
\langle eT^\mu\,_\mu \rangle_{\rm Reg} = \lim_{x \rightarrow y}\{
\frac{1} {2}(1-\gamma^5)\frac{1}{2}(f -\sigma^3)\delta(x-y)\} +
2i\langle \partial_\mu J^\mu_L \rangle_{\rm Reg}
\end{equation}
connecting the ABJ and conformal anomalies\cite{cpt}.

%%%!!!
The upshot is that the coupling of $C_\mu$ to an anomalous
current poses no consistency problems in the quantum theory.  In
particular, the presence of the mass term $C_\mu C^\mu$ is
compatible with all local invariances in the action. Its Green
functions can and does satisfy modified Ward identities with
additional terms that imply non-transversality.
%%The fact that the current coupled to $C_\mu$ is anomalous
%%does not mean that Lorentz invariance has been sacrificed.
%%There is an incompatibility between maintaining local Lorentz and
%%invariance under local singlet axial and scale transformations.
%%And this occurs despite the superficial similarity between
%%the torsion coupling and an Abelian interaction.
%%there are no Lorentz anomalies in four dimensions \cite{Nieh2},
%%even in the presence of torsion \cite{Kimura}.
%%This is compatible with the fact that the
%%regularization of Ref.\cite{invariant} employed here explicitly
%%maintains all the gauged symmetries, including local Lorentz invariance.
%%What is true is that Lorentz symmetry cannot be preserved at the
%% quantum level if we also insist on simultaneously maintaining
%%the Abelian {\it gauge} symmetry.
%%We may however keep Lorentz invariance but tolerate anomalies
%%in currents not coupled to gauge fields. This is precisely the case
%%here for Lorentz invariance is explicit, while Abelian $\gamma^5$
%%rotations and scaling symmetry are anomalously broken by
%%the massive regulators and are not gauged as local symmetries.
With regard to the Abelian part of gauged internal symmetries, we
still require ${\rm Tr}(T^a)$ to vanish for the regularization to
work in curved space \cite{invariant}.
%%%!!!

%%%!!!
The appearance of a mass term for $C_\mu$ does have one important
consequence.
%%%!!!
The combination $C_\mu$ is complex, and the counterterms of the
Lorentzian signature Lagrangian required by the vacuum
polarization diagrams are of the form
\begin{eqnarray}
eg^{\mu\alpha}g^{\nu\beta}(\partial_\mu C_\nu -\partial_\nu C_\mu)
(\partial_\alpha C_\beta -\partial_\beta C_\alpha) =
eg^{\mu\alpha}g^{\nu\beta}&&[-(\partial_\mu B_\nu -\partial_\nu
B_\mu) (\partial_\alpha B_\beta -\partial_\beta B_\alpha)\cr
\nonumber\\
&&+ 2i(\partial_\mu B_\nu -\partial_\nu B_\mu) (\partial_\alpha
{A}_\beta -\partial_\beta{A}_\alpha)\cr
\nonumber\\
&&+ (\partial_\mu {A}_\nu -\partial_\nu {A}_\mu) (\partial_\alpha
{A}_\beta -\partial_\beta{A}_\alpha)],
\end{eqnarray}
and
\begin{equation}
eg^{\mu\nu}C_\mu C_\nu = eg^{\mu\nu}(-B_\mu B_\nu + 2iB_\mu{A}_\nu
+ {A}_\mu{A}_\nu).
\end{equation}
Therefore the anti-hermitian cross terms are Lorentz invariant,
but CPT-odd. In a related work \cite{cpt} it was pointed out that
a truly Weyl theory which preserves the chirality of the
gravitational interaction should violate CPT through $B_\mu$
effects because of the non-hermitian coupling; and that these
signatures of chiral gravity \cite{Ash} will already be manifest
at the level of quantum field theory in curved spaces in Weyl
fermion loops processes with external chiral gravitational fields.
An example is precisely the vacuum polarization diagram with two
complex left-handed spin connection vertices. The flat vierbein
limit with nontrivial $C_\mu$ evaluated here indeed confirms the
presence of these CPT-violating terms in the effective action.

In general, we may decompose the torsion as $T_{\mu\nu} = {2
\over 3}(B_\mu e_{\nu A} - B_\nu e_{\mu A})
             + {1\over 3}\epsilon_{\mu\nu A\alpha}\tilde{A^\alpha}
             + Q_{\mu\nu A}$,
where $Q_{\mu\nu A} $ is constrained by $E^{\nu A} Q_{\mu\nu A} =
\epsilon^{\mu\nu A \alpha} Q_{\mu\nu A}=0.$ Since $Q_{\mu\nu A}$
does not couple to fermions, the spin connection in the fermion
coupling can be restricted to
\begin{equation}
A_{\mu AB} = \omega_{\mu AB} + {2 \over 3}(B_\nu E^\nu _A e_{\mu
B} - B_\nu E^\nu_B e_{\mu A}) - {1\over
6}{\epsilon}_{AB\mu\alpha}\tilde{A^\alpha}
\end{equation}
As a result, the Samuel-Jacobson-Smolin action\cite{Ash} (which
is the (anti)-self-dual part of the Einstein-Hilbert-Palatini
action), ${i\over{8\pi G}}\int e^A \wedge e^B \wedge {1\over
2}(i*+1)F_{AB}$, is equivalent to ${1\over {16\pi G}}\int d^4x
(eR - 4i\partial_\mu (eC^\mu) +{8\over 3} C_\mu C^\mu)$. Modulo
the total divergence term which is not reproduced by perturbation
theory, the $C_\mu C^\mu$ counter term required by vacuum
polarization computations is of the same form as the
Samuel-Jacobson-Smolin action in the teleparallel limit.
%Barring a bare mass,
%we can therefore expect the relevant mass term for $C_\mu$ in the
%renormalized effective action of Weyl fermions coupled to full-fledged
%background gravitational fields to be the corresponding quadratic term
%of the full Samuel-Jacobson-Smolin action; and the coefficient of
%$C_\mu C^\mu$ in the effective action is
%thus ${1 \over {(6\pi G_{renor.})}}$, implying a Planck scale mass.

Phenomenological consequences of massive axial torsion modes have
been discussed before\cite{Shap}. In our present context, we wish
to pursue the extent to which chirality can be used as a defining
characteristic of particle interactions, including gravity.  That
context required us to use Weyl spinors, and also the Ashtekar
formulation of gravity. A net consequence is that both axial
torsion as well as vector torsion trace are needed, but with a
relative phase which ruins CPT invariance because of the chiral
nature of the fields. Massive modes appear as consequences of the
anomalous non-conservation of the current to which these torsion
modes are coupled. Our results to some degree extend those of
Ref. \cite{Shap} to cover the case of torsion trace as well. At
low energies compared to the torsion mass, the fermion-torsion
interaction produces a four-fermion coupling. Present high energy
experimental data on four-fermion vertices sets the lower bound
for torsion mass at above roughly 200GeV \cite{Shap}.

The question of mass generation via anomalies has had a storied
past. In the Schwinger model in 2D, the physical degree of
freedom of the photon is equivalent to a free massive
boson\cite{Schwinger,Brown} since the interaction term can be
transformed away by an axial rotation. The mass of the boson
stems from the ABJ anomaly, which gives rise to an infra-red pole
in the polarization tensor. The value of the mass is uniquely
determined when the vacuum polarization tensor is regularized to
satisfy vector gauge invariance at the expense of axial vector
current conservation. On the other hand, the chiral Schwinger
model in 2D is anomalous albeit exactly solvable. The resultant
photon mass, while still finite, carries an ambiguity as the
previous condition of gauge invariance is now absent. It can be
made to vanish, while preserving the (V-A) form for the coupling,
but we then lose unitarity \cite{Jackiw}.
%%%!!!
This paper discusses the corresponding chiral situation in 4D,
but without loss of any local gauge invariance. By retaining
explicitly all local gauge symmetries and the holomorphic
dependence on the left-handed spin connection in the
regularization, we end up with a vacuum polarization tensor that
is non-transverse, and gives a mass to $C_\mu =iB_\mu + A_\mu$.
These torsion modes are massive because of ABJ and scaling
anomalies, with generated masses naturally of the order of the
regulator scale. Since these are the only modes that can couple
to spin $\frac{1}{2}$ fermions, large regulator masses, or high
cut-off scales in the context of effective field theories,
naturally explain why torsion and its associated effects,
including CPT violations from $B_\mu$ couplings, have so far
escaped detection.
%%%!!!

\acknowledgments

The research for this work has been supported in part by funds
from the U.S. Department of Energy under Grant No.
DE-FG05-92ER40709, the National Center for Theoretical Sciences,
Taiwan, and the National Science Council of Taiwan under Grant
No. NSC 89-2112-M-0060-050.

\appendix
\section{Notations and some relevant identities}

Lorentz indices are denoted by uppercase Latin letters while
Greek symbols are spacetime indices. The flat Lorentzian metric
is $\eta_{AB}= {\rm diag}(-1,1,1,1)$, and $g_{\mu\nu} = \eta_{AB}
e^A_\mu e^B_\nu$.

Let us recall from the definition of torsion
\begin{equation}
T_A = \frac{1}{2}T_{A\mu\nu}dx^\mu\wedge{dx^\nu} = de_A +
A_{AB}\wedge e^B,
\end{equation}
that the general solution for invertible vierbein is
\begin{equation}
 A_{\mu AB} = \omega_{\mu AB} - \frac{1}{2}[T_{A\sigma\mu}E^\sigma_B -
T_{B\sigma\mu}E^\sigma_A - T_{C\rho\sigma}e_\mu^C E^\rho_A
E^\sigma_B ],
\end{equation}
with $E^{\mu A}$ being the inverse of the vierbein $e_{\mu A}$;
and $\omega_{AB}$ is the torsionless spin connection ($ de_A +
\omega_{AB}\wedge  e^B =0$) which can be solved as
\begin{equation}
\omega_{\mu AB} = \frac{1}{2} [E^\nu_A (\partial_\mu e_{\nu B}
-\partial_\nu e_{\mu B}) - E^\nu_B (\partial_\mu e_{\nu A}
-\partial_\nu e_{\mu A}) -E^\alpha_A E^\beta_B (\partial_\alpha
e_{\beta C} -\partial_\beta e_{\alpha C})e^C_\mu ].
\end{equation}
Spin $\frac{1}{2}$ fermions couple minimally to torsion through
the spin connection $\frac{i}{2}A_{\mu AB}\sigma^{AB},$
$(\sigma^{AB} = \frac{1}{4}[\gamma^A, \gamma^B])$, in the Dirac
operator $i{D\kern-0.15em\raise0.17ex\llap{/}\kern0.15em\relax} =
\gamma^\mu(i\partial_\mu + \frac{i}{2}A_{\mu AB}\sigma^{AB} +
W_{\mu a}{T}^{a})$. Here $W_{\mu a}$ denotes the generic internal
gauge connection in the ${T}^a$ representation. By substituting
for $A_{\mu AB}$ the interaction reduces to
\begin{equation}
e\frac{i}{2}{\overline{\Psi}}
{A\kern-0.15em\raise0.17ex\llap{/}\kern0.15em\relax}_{AB}\sigma^{AB}\Psi
= \frac{1}{2}(-i \omega_{\mu AB}J^\nu -
\frac{1}{2}\epsilon_{AB}\,^{CD} \omega_{\mu CD}J^{5\nu}) E^{\mu
B} e^A_\nu
 -(iB_\mu J^\mu - \frac{1}{4e}\tilde{A_\mu}J^{5\mu})
\end{equation}
where $B_\mu \equiv \frac{1}{2} T_{A\mu\nu}E^{\nu A}$ and
$\tilde{A}_\mu \equiv \frac{1}{2}g_{\mu\lambda}
{\tilde\epsilon}^{\lambda\nu\alpha\beta}e_{\nu A}
T^A_{\alpha\beta} $ are precisely the trace and axial parts of the
torsion, while $J^\mu = \overline{\Psi}e\gamma^\mu\Psi$ and
$J^{5\mu} = \overline{\Psi}e\gamma^\mu\gamma^5\Psi$ are the
respective vector and axial-vector currents. Note that only the
axial torsion and torsion trace interactions are present. In
particular, for chiral fermions,
\begin{eqnarray}
e\frac{i}{2}{\overline{\Psi}}_{L,R}
{A\kern-0.15em\raise0.17ex\llap{/}\kern0.15em\relax}_{AB}
\sigma^{AB}\Psi_{L,R}
 &=& \frac{1}{2}(-iA_{\mu AB} \pm \frac{1}{2}\epsilon_{AB}\,^{CD}
A_{\mu CD}) E^{\mu B} e^A_\nu J^\nu_{L,R}\cr
\nonumber\\
&=&\frac{1}{2}(-i\omega_{\mu AB} \pm
\frac{1}{2}\epsilon_{AB}\,^{CD} \omega_{\mu CD}) E^{\mu B}
e^A_\nu J^\nu_{L,R}
 -(iB_\mu \pm \frac{1}{4e}\tilde{A_\mu})J^\mu_{L,R}
\end{eqnarray}
with $J^\mu_{L,R} ={\overline{\Psi}}_{L,R}e\gamma^\mu \Psi_{L,R}
= \mp J^{5\mu}_{L,R}$. This shows that left(right)-handed chiral
fermions couple to the left(right)-handed (or
anti-self-dual(self-dual)) projection of the spin connection, and
chiral fermions interact only with the $(iB_\mu \pm
\frac{1}{4e}{\tilde A})$ components.

\section{Vacuum Polarization tensor of internal gauge fields}

We begin by observing that the gauge current $ J^{\mu a}_L =
-{\overline\Psi}_L e\gamma^\mu T^a \Psi_L $ is regularized as
\begin{equation}
\langle J^{\mu a}_L(x)\rangle_{\rm Reg} =
%-\lim_{x \rightarrow y}{\rm Tr}\{\gamma^\mu {\cal T}^a
\frac{1}{2}(1-\gamma^5)
%{1\over {i{D\kern-0.15em\raise0.17ex\llap{/}\kern0.15em\relax}}}
%\frac{1}{2}(f -\sigma^3)\delta(x-y)\}\cr
%\nonumber\\
-\lim_{x \rightarrow y}{\rm Tr}\{\gamma^\mu {\cal T}^a
\frac{1}{2}(1-\gamma^5) {1\over
{i{D\kern-0.15em\raise0.17ex\llap{/}\kern0.15em\relax}}}
\frac{1}{2}(\sum_n
C_n{{i{D\kern-0.15em\raise0.17ex\llap{/}\kern0.15em\relax}
(i{D\kern-0.15em\raise0.17ex\llap{/}\kern0.15em\relax})^\dagger}\over
{[i{D\kern-0.15em\raise0.17ex\llap{/}\kern0.15em\relax}
{(i{D\kern-0.15em\raise0.17ex\llap{/}\kern0.15em\relax})^\dagger}
+ m^2_n]}}
 -\sigma^3)\delta(x-y)\}.
\end{equation}
For pure internal gauge fields in flat spacetime we set ${\tilde
A}_\mu = B_\mu = 0$ in the Dirac operator to yield
\begin{equation}
i{D\kern-0.15em\raise0.17ex\llap{/}\kern0.15em\relax} =
(i{D\kern-0.15em\raise0.17ex\llap{/}\kern0.15em\relax})^\dagger =
i{\partial\kern-0.15em\raise0.17ex\llap{/}\kern0.15em\relax} +
{W\kern-0.15em\raise0.17ex\llap{/}\kern0.15em\relax}_a{\cal T}^a .
\end{equation}
The positive-definite operator which appears in the regulator
function is
\begin{equation}
i{D\kern-0.15em\raise0.17ex\llap{/}\kern0.15em\relax}
(i{D\kern-0.15em\raise0.17ex\llap{/}\kern0.15em\relax})^\dagger =
-\Box +
i{\partial\kern-0.15em\raise0.17ex\llap{/}\kern0.15em\relax}
{W\kern-0.15em\raise0.17ex\llap{/}\kern0.15em\relax}_a{\cal T}^a
+{W\kern-0.15em\raise0.17ex\llap{/}\kern0.15em\relax}_a{\cal T}^a
i{\partial\kern-0.15em\raise0.17ex\llap{/}\kern0.15em\relax} +
{W\kern-0.15em\raise0.17ex\llap{/}\kern0.15em\relax}_a{\cal T}^a
{W\kern-0.15em\raise0.17ex\llap{/}\kern0.15em\relax}_b{\cal T}^b
\end{equation}
where $\Box =\partial_\mu \partial^\mu$. Thus the current has the
expansion
\begin{eqnarray}
\langle J^{\mu a}_L(x)\rangle_{\rm Reg}&=& \lim_{x \rightarrow
y}\frac{1}{2}{\rm Tr}\{{\cal T}^a{\cal T}^b\} {\rm
Tr}\{\gamma^\mu \frac{1}{2}(1-\gamma^5)\sum_n C_n [...-
{{m_n}\over{(-\Box + m^2_n)}}
{W\kern-0.15em\raise0.17ex\llap{/}\kern0.15em\relax}_b
{m_n\over{(-\Box + m^2_n)}}\cr
\nonumber\\
&&+{{i{\partial\kern-0.15em\raise0.17ex\llap{/}\kern0.15em\relax}}
\over{(-\Box+
m^2_n)}}{W\kern-0.15em\raise0.17ex\llap{/}\kern0.15em\relax}_b
{{i{\partial\kern-0.15em\raise0.17ex\llap{/}\kern0.15em\relax}}
\over{(-\Box + m^2_n)}}+...]\delta(x-y)\}. \label{B4}
\end{eqnarray}
In the above, we display only terms which will contribute to the
vacuum polarization tensor i.e. terms which are linear in $W_{\mu
a}$ since the vacuum polarization $\Pi^{\mu \nu a b} $ is defined
through the Fourier transform
\begin{eqnarray}
{1\over{(2\pi^4)}}\int d^4k \, \Pi^{\mu\nu ab}{e^{ik.(x-y)}} &
\equiv &\left.{{\delta \langle J^{\mu a}(x)\rangle} \over{\delta
W_{\nu b}(y)}}\right|_{W =0} \cr
\nonumber\\
&=&\lim_{x \rightarrow z}\frac{1}{2}{\rm Tr}\{{\cal T}^a{\cal
T}^b\} {\rm Tr}\{\gamma^\mu\frac{1}{2}(1-\gamma^5)\sum_n C_n [
-{{m^2_n}\over{(-\Box + m^2_n)}}\gamma^\nu\delta(x-y)\cr
\nonumber\\
&&+{{i{\partial\kern-0.15em\raise0.17ex\llap{/}\kern0.15em\relax}}
\over{(-\Box + m^2_n)}}\gamma^\nu \delta(x-y)
{i{\partial\kern-0.15em\raise0.17ex\llap{/}\kern0.15em\relax}}\,]
[{1\over{(-\Box + m^2_n)}}\delta(x-z)]\}.
\end{eqnarray}
Denoting the Dirac delta functions by
\begin{equation}
\delta(x-y) = {1\over{(2\pi)^4}}\int d^4k\,e^{ik.(x-y)}, \qquad
\delta(x-z) = {1\over{(2\pi)^4}}\int d^4p\,e^{ip.(x-z)},
\end{equation}
and bearing in mind that $\frac{1}{2}{\rm Tr}\{{\cal T}^a{\cal
T}^b\} = {\rm Tr}\{{T}^a{T}^b\}$, the vacuum polarization tensor
is therefore
\begin{eqnarray}
\Pi^{\mu\nu ab}={\rm Tr}\{T^a T^b\}\sum_n C_n \int {d^4p\over
{(2\pi)^4}} \, {\rm Tr}\{\gamma^\mu \frac{1}{2}(1-\gamma^5) [&&-
{{m_n}\over{[(p+k)^2 + m^2_n]}}\gamma^\nu{m_n\over{(p^2 +
m^2_n)}}\cr
\nonumber\\
&&+{{({p\kern-0.15em\raise0.17ex\llap{/}\kern0.15em\relax} +
{k\kern-0.15em\raise0.17ex\llap{/}\kern0.15em\relax})}
\over{[(p+k)^2 + m^2_n]}}\gamma^\nu
{{{p\kern-0.15em\raise0.17ex\llap{/}\kern0.15em\relax}}
\over{(p^2+ m^2_n)}}]\}. \label{B7}
\end{eqnarray}
By using the Feynman parametrization
\begin{equation}
{1\over{[(p+k)^2 + m^2_n](p^2 + m^2_n)}} = \int^1_0
{{dz}\over{[p^2 + 2kp(1-z) + k^2(1-z) + m^2_n]^2}} =\int^1_0
{{dz}\over{[(p')^2 + k^2z(1-z) + m^2_n]^2}},
\end{equation}
with ${p'} \equiv p + k(1-z)$, the first term of Eq.\ (\ref{B7})
can be rewritten as
\begin{eqnarray}
\int {{d^4p}\over {(2\pi)^4}}I^{\mu \nu n}_1 &\equiv& -\int
{{d^4p}\over {(2\pi)^4}}{\rm Tr}\{\gamma^\mu
\frac{1}{2}(1-\gamma^5) [{{m_n}\over{[(p+k)^2 +
m^2_n]}}\gamma^\nu{m_n\over{(p^2 + m^2_n)}}\cr
\nonumber\\
&=&-2g^{\mu\nu}\int {{d^4p}\over {(2\pi)^4}} \int^1_0 dz \,
{{m^2_n}\over{[p^2 + k^2z(1-z) + m^2_n]^2}} \,\, ;
\end{eqnarray}
while the final contribution to $\Pi^{\mu \nu a b}$ of Eq.\
(\ref{B7}) can be reexpressed as
\begin{eqnarray}
\int {{d^4p}\over {(2\pi)^4}} I^{\mu\nu n}_2 &\equiv& \int
{{d^4p}\over {(2\pi)^4}}{\rm Tr}\{\gamma^\mu
\frac{1}{2}(1-\gamma^5)
{{({p\kern-0.15em\raise0.17ex\llap{/}\kern0.15em\relax} +
{k\kern-0.15em\raise0.17ex\llap{/}\kern0.15em\relax})} \over
{[(p+k)^2 + m^2_n]}}\gamma^\nu
{{{p\kern-0.15em\raise0.17ex\llap{/}\kern0.15em\relax}}
\over{(p^2 + m^2_n)}}\}\cr
\nonumber\\
%&=&2\int^1_0{dz}\int{{d^4p'}\over {(2\pi)^4}}
%{{[p'+ kz]_\alpha[p'- k(1-z)]_\beta[g^{\mu\alpha}g^{\nu\beta}
%-g^{\mu\nu}g^{\alpha\beta} + g^{\mu\beta}g^{\alpha\nu}}]
%\over{[{p'}^2 + k^2z(1-z) + m^2_n]^2}}\cr
%\nonumber\\
&=&2\int^1_0{dz}\int{{d^4p}\over {(2\pi)^4}} {{-2z(1-z)(k^\mu
k^\nu - k^2g^{\mu\nu}) - g^{\mu\nu}[(p^2/2) + k^2z(1-z)]}
\over{[p^2 +  k^2z(1-z) + m^2_n]^2}}.
\end{eqnarray}
In total,
\begin{equation}
\Pi^{\mu\nu ab}={\rm Tr}\{T^a T^b\}\sum_n C_n \int {{d^4p}\over
{(2\pi)^4}} \, ( I^{\mu \nu n}_1 + I^{\mu \nu n}_2 ), \label{B11}
\end{equation}
with $C_n = (-1)^n$ and $m_n = n\Lambda$. For the explicit
computation of the polarization tensor, a few identities related
to our regulator function $f(p^2)$ are required. We have
\begin{equation}
{\tilde f}(p^2) \equiv \sum^{\infty}_{n= -\infty}
{{(-1)^n}\over{(p^2 + n^2)}} ={\pi
\over{\sqrt{p^2}\sinh(\pi\sqrt{p^2})}} ={{f(p^2)}\over p^2},
\end{equation}
\begin{equation}
{{d{\tilde f}}\over {dp^2}} = -\sum^{\infty}_{n=-\infty}
{{(-1)^n}\over{[p^2 + n^2]^2}},
%=-{\pi\over{2p^3\sinh(\pi p)}}-{{\pi^2\coth(\pi p)}\over {2p^2\sinh(\pi
%p)}}
\end{equation}
and
\begin{equation}
\sum^{\infty}_{n=-\infty} {{(-1)^n n^2}\over{[p^2 + n^2]^2}}
={\tilde f}(p^2) + p^2{{d{\tilde f}(p^2)}\over {d p^2}} ={ d\over
{dp^2}} [p^2{\tilde f}(p^2)]
\end{equation}

By writing $\Pi^{\mu\nu a b}$ as a sum of transverse and
non-transverse parts,
\begin{equation}
\Pi^{\mu\nu ab}={\rm Tr}\{T^a T^b\}( A^{\mu \nu}_T + A^{\mu
\nu}_{NT} ),
\end{equation}
we observe that the non-transverse contribution vanishes
identically since
\begin{eqnarray}
A^{\mu\nu}_{NT} &=& -2g^{\mu\nu}\int^1_0 dz \int{{d^4p}\over
{(2\pi)^4}} \sum_n C_n {{[k^2z(1-z) + (p^2/2) +m^2_n]}\over{ [p^2
+ k^2z(1-z) + m^2_n]^2}}\cr
\nonumber\\
&=& 2g^{\mu\nu}\int^1_0 dz \int{{d^4p}\over {(2\pi)^4}} \sum_n
(-1)^n\left(-{{1}\over{[p^2 + k^2z(1-z) + n^2\Lambda^2]}}
+{{p^2}\over{2[p^2 + k^2z(1-z) + n^2\Lambda^2]^2}}\right)\cr
\nonumber\\
%&=& 2g^{\mu\nu}\Lambda^2\int^1_0 dz
%\int^\infty_0 {{dp^2 \,\pi^2 p^2}\over {(2\pi)^4}}
%[-{\tilde f}(p^2+s) -\frac{1}{2}p^2{\partial\over
%{\partial p^2}}{\tilde f}(p^2 + s)] \cr
%\nonumber\\
&=& -g^{\mu\nu}\Lambda^2\int^1_0 dz {1
\over{(16\pi^2)}}\int^\infty_0 dp^2 \, {\partial\over{\partial
p^2}}[p^4{\tilde f}(p^2 + s)] = 0.
%\cr
%\nonumber\\
%&=& -{{g^{\mu\nu}\Lambda^2}\over{16\pi^2}}
%\int^1_0 dz \left.[p^4{\tilde f}(p^2 + s)]\right|^\infty_{p^2=0}
%=0.
\end{eqnarray}
In the above, we have labeled $s={{k^2 z(1-z)}\over {\Lambda^2}}$
for convenience. The transverse piece is
\begin{eqnarray}
A^{\mu\nu}_T &=& 2\int^1_0{dz}\int{{d^4p}\over {(2\pi)^4}}\sum_n
(-1)^n {{-2z(1-z)(k^\mu k^\nu - k^2g^{\mu\nu})}
\over{\Lambda^4[p^2/\Lambda^2 +  k^2z(1-z)/\Lambda^2 + n^2]^2}}\cr
\nonumber\\
&=& {1 \over {(4\pi^2)}}(k^\mu k^\nu - g^{\mu\nu}k^2) \int^1_0
dz\, z(1-z) \int^\infty_0 dp^2 \, p^2 {\partial \over {\partial
p}} {\tilde f}(p^2 + s)\cr
\nonumber\\
&=& {1\over {(4\pi^2)}}(k^\mu k^\nu - g^{\mu\nu}k^2) \int^1_0
dz\,2z(1-z)\ln \tanh({\pi\over {2}}\sqrt{s}), \label{B17}
\end{eqnarray}
on rescaling $p/\Lambda \rightarrow p$ in the intermediate step,
and noting that
\begin{eqnarray}
\int^\infty_0 dp^2 \, p^2 {\partial\over {\partial p^2}} {\tilde
f}(p^2 + s) &=& \left. p^2{\tilde f}(p^2+s)\right|^\infty_{p^2=0}
-\int^\infty_0 dp^2 \, {\tilde f}(p^2 + s)\cr
\nonumber\\
&=& -\int^\infty_s {{\pi d(p^2+s)}\over {\sqrt{p^2+s}\sinh(\pi
\sqrt{p^2+s})}} \cr
\nonumber\\
&=&-2\left.\ln \tanh({{\pi x}\over{2}})\right|^{\infty}_{x
=\sqrt{s}}.
\end{eqnarray}
We therefore arrive at
\begin{eqnarray}
\Pi^{\mu\nu ab}&=&{\rm Tr}\{T^a T^b\}A^{\mu\nu}_T \cr
\nonumber\\
&=&{{{\rm Tr}\{T^a T^b\}}\over {24\pi^2}} (k^\mu k^\nu -
g^{\mu\nu}k^2)[\ln ({{k^2}\over{\Lambda^2}}) - \frac{5}{3} + 2
\ln ({\pi \over 2})] \label{B19}
\end{eqnarray}
after integration over $z$. This agrees with result obtained
previously in Refs. \cite{Aoki} and \cite{Okuyama} using the
generalized Pauli-Villars scheme with doubling in external
left-right fermionic space (rather than the method here of
doubling in the internal $T^a$ space).

The vacuum polarization tensor for internal gauge fields is
therefore still transverse, and internal gauge fields remain
massless. This is expected since the regularization explicitly
respects the local internal symmetries gauged by $W_{\mu a}$.

%---------------------------------------------


\begin{references}
\bibitem[*]{byline1}  Electronic Address: laynam@vt.edu.
\bibitem[\dagger]{byline2}Electronic Address: cpsoo@phys.nthu.edu.tw

\bibitem{ABJ} S. L. Adler, Phys. Rev. {\bf 177} (1969) 2426; J. S. Bell
and R. Jackiw, Nuovo Cimento {\bf 60A} (1969) 47.
\bibitem{Yajima} S. Yajima and T. Kimura, Prog. Theo. Phys. {\bf 74}
(1985) 866; S. Yajima, Prog. Theo. Phys. {\bf 79} (1988) 535;
Class. Quantum Grav. {\bf 5} (1988) L207; {\it ibid.} {\bf 13}
(1996) 2423; {\bf 14} (1997) 2853.
\bibitem{Chandia} O. Chandia and J. Zanelli, Phys. Rev. {\bf D55}
(1997) 7580; {\it ibid.} {\bf D58} (1998) 045014; hep-th/9708138;
Y. N. Obukhov, E. W. Mielke, J. Budzies and F. W. Hehl, Found.
Phys. {\bf 27} (1997) 1221.
\bibitem{cpabj} C. Soo, Phys. Rev. {\bf D59} (1999) 045006.
\bibitem{Nieh-Yan} H. T. Nieh and M. L. Yan, J. Math. Phys. {\bf 23}(3)
(1982) 373.
\bibitem{Ash}A.\ Ashtekar, Phys.\ Rev.\ Lett. {\bf 57} (1986) 2244;
Phys.\ Rev. {\bf D36} (1986) 1587; {\it New perspectives in
canonical gravity}, (Bibliopolis, Naples, 1988);  {\it Lectures
on nonperturbative canonical gravity}, (World Scientific,
Singapore, 1991); J.\ Samuel, Pram$\bar{\rm a}$na J.\ Phys. {\bf
28} (1987) L429; Class.\ Quantum Grav. {\bf 5} (1988) L123; T.\
Jacobson and L.\ Smolin, Phys. Lett. {\bf B196} (1987) 39, Class.
Quantum Grav. {\bf 5} (1988) 583.
\bibitem{cps} L. N. Chang and C. Soo, Phy. Rev. {\bf D53} (1996) 5682.
\bibitem{cpt} C. Soo and L. N. Chang, hep-th/9702171.
\bibitem{invariant} L. N. Chang and C. Soo, Phys. Rev. {\bf D55} (1997)
2410.
\bibitem{Aoki} S. Aoki and Y. Kikukawa, Mod. Phys.
Lett. {\bf A37} (1993) 3517.
\bibitem{Okuyama} K. Okuyama and H. Suzuki,
hep-th/9603062; Phys. Lett. {\bf B382} (1996) 117 and references
therein for Pauli-Villars regularization with doubling in
external left-right Lorentz space.
\bibitem{PeskinBook} See, for instance, M. E. Peskin and D. V.
Schroeder, {\it An introduction to quantum field theory}, Section
7.5 of Chapter 7, (Addison-Wesley, 1995).
\bibitem{DeWitt} See, for instance, B. S. DeWitt, {\it Dynamical theory
of groups and fields}, (Gordon and Breach, New York, 1965).
\bibitem{Slavnov} S. A. Frolov and A. A. Slavnov, Phys. Lett. {\bf B309}
(1993) 344; Nucl. Phys. {\bf B411} (1994) 647.
%\bibitem{Nieh2} H.T. Nieh and M. L. Yan, Ann. Phys. {\bf 138}, 237
%(1982).
%\bibitem{Nieh2} L. Alvarez-Gaume and E. Witten,
%Nucl. Phys. {\bf B234} (1984) 269; L. N. Chang and H. T. Nieh, Phys.
%Rev. Lett. {\bf 53} (1984) 21.
%H. T. Nieh, Phys. Rev. Lett. {\bf 53} (1984) 2219;
%\bibitem{Kimura} S. Yajima and T. Kimura, Phys. Lett. {\bf B173} (1986)
%154.
\bibitem{Shap} S. M. Carroll and G. B. Field, Phys. Rev {\bf D50} (1994)
3867; A. S. Belyaev and I. L. Shapiro, Nucl. Phys. {\bf B543}
(1999) 20 and references therein.
\bibitem{Schwinger} J. Schwinger, Phys. Rev. {\bf 128} (1962) 2425.
\bibitem{Brown} D. G. Boulware and W. Gilbert, Phys. Rev. {\bf 126}
(1962) 1563; L. S. Brown, Nuovo, {\bf 29} (1963) 617; B. Zumino,
Phys. Lett. {\bf 10} (1964) 224.
\bibitem{Jackiw} R. Jackiw and R. Rajaraman, Phys. Rev. Lett. {\bf 54}
(1985) 1219; A. Das. Phys. Rev. Lett {\bf 55} (1985) 2126.



\end{references}
\end{document}